\documentclass[conference]{IEEEtran}
\IEEEoverridecommandlockouts
\usepackage{cite}
\usepackage{amsmath,amssymb,amsfonts}
\usepackage{algorithmic}
\usepackage{algorithm}
\usepackage{graphicx}
\usepackage[caption=false,font=footnotesize]{subfig}
\usepackage{textcomp}
\usepackage[dvipsnames,svgnames,table]{xcolor}
\usepackage{booktabs}
\usepackage{makecell}
\newcommand{\methodname}{TIDE}
\newcommand{\method}{\textbf{\texttt{\methodname}}}
\newcommand{\gray}[1]{\textcolor{gray}{#1}}
\newcommand{\venue}[1]{{\scriptsize\textcolor{gray!70!black}{[#1]}}}
\colorlet{bestcell}{red!13}
\colorlet{secondcell}{blue!10}
\newcommand{\best}[1]{\cellcolor{bestcell}\textbf{#1}}
\newcommand{\second}[1]{\cellcolor{secondcell}\textbf{#1}}
\DeclareRobustCommand{\swatch}[1]{{\setlength{\fboxsep}{0pt}\setlength{\fboxrule}{0.4pt}\raisebox{-0.2ex}{\fcolorbox{gray!80}{#1}{\rule{0pt}{1.8ex}\hspace{5ex}}}}}
\newcommand{\band}[1]{\gray{\textit{#1}}}
\definecolor{hdrA}{HTML}{8A1F3D}
\definecolor{hdrB}{HTML}{16479B}
\definecolor{hdrC}{HTML}{15683A}
\definecolor{hdrD}{HTML}{4D4D4D}
\newcommand{\hd}[2]{\cellcolor{#1}\textcolor{white}{\textbf{#2}}}
\definecolor{oursname}{HTML}{FBE5D0}

\newcommand{\std}[1]{{\scriptsize\textcolor{ForestGreen}{($\pm$#1)}}}
\usepackage{tcolorbox}
\usepackage{ntheorem}
\theoremseparator{:}

\theoremstyle{nonumberplain}
\theorembodyfont{\normalfont}

\usepackage{hyperref}
\usepackage{balance}
\def\BibTeX{{\rm B\kern-.05em{\sc i\kern-.025em b}\kern-.08em
    T\kern-.1667em\lower.7ex\hbox{E}\kern-.125emX}}
\begin{document}

\title{Pod-wise Multimodal Federated Learning for Heterogeneous IoT}

\author{
\IEEEauthorblockN{Joy Okolo, Zihao Ding, and Jun Huang}
\IEEEauthorblockA{Department of Electrical Engineering and Computer Science, South Dakota State University, Brookings, SD 57007, USA\\
Email: \{joy.okolo, zihao.ding\}@jacks.sdstate.edu, jun.huang@sdstate.edu}
}

\maketitle

\begin{abstract}
Federated learning (FL) is well suited to Internet of Things (IoT) sensing. However, devices carry different sensors and come online in periodic patterns, and as such, the sensors among the participants change from round to round. Under naive averaging, devices without a sensor overwrite its parameters, and a single-sensor device never learns how its sensor relates to the others. Existing methods either redesign local training for mismatched sensors (assuming always-available devices) or select clients without regard to their sensors. We propose \underline{T}ier-aware \underline{I}ntra-pod \underline{D}ecoupled \underline{E}ncoders (\method{}). Specifically, each sensor has its own encoder and auxiliary head, whose parameters are merged only within the ``pod'' (i.e., the devices that hold the sensor), while the fusion layer and the task head remain shared by all devices. On three multimodal activity recognition datasets against eight baselines, \method{} reaches 78.06\% on ActionSense, 7.63 points above the next method, is the most accurate on all three datasets at half its own communication, and at equal communication leads on ActionSense and is ahead within seed variation on the other two.
\end{abstract}


\section{Introduction}
\label{sec:intro}

Federated learning (FL) allows many devices to train one shared model without sending their private data to a server~\cite{McMahan2016CommunicationEfficientLO,Wu2026TNSE,Che2026arXiv,Ding2026ICDCS,Ding2026MNET,Pudasaini2026HPSRAgri,Ding2026arXivMUTE}. As such, it is a natural fit for the Internet of Things (IoT), where wearables and mobile devices continuously collect private data through several sensors at once~\cite{zhao2022multimodal,Wu2026COMST,Wu2026Network,Wu2026TON,Ding2025IPCCC,Ding2026TAAS,Ding2026ICNC,Huang2020TWC,Huang2019WCM,Che2027PAKDD,Wu2026arXivRouting,Wu2026arXivREBASE,Wu2026arXivCrystalMem}. As an example, a smartwatch may carry an accelerometer and a gyroscope, while a patch worn next to it carries only one of them.

\begin{figure}[t]
  \centering
  \includegraphics[width=1\columnwidth]{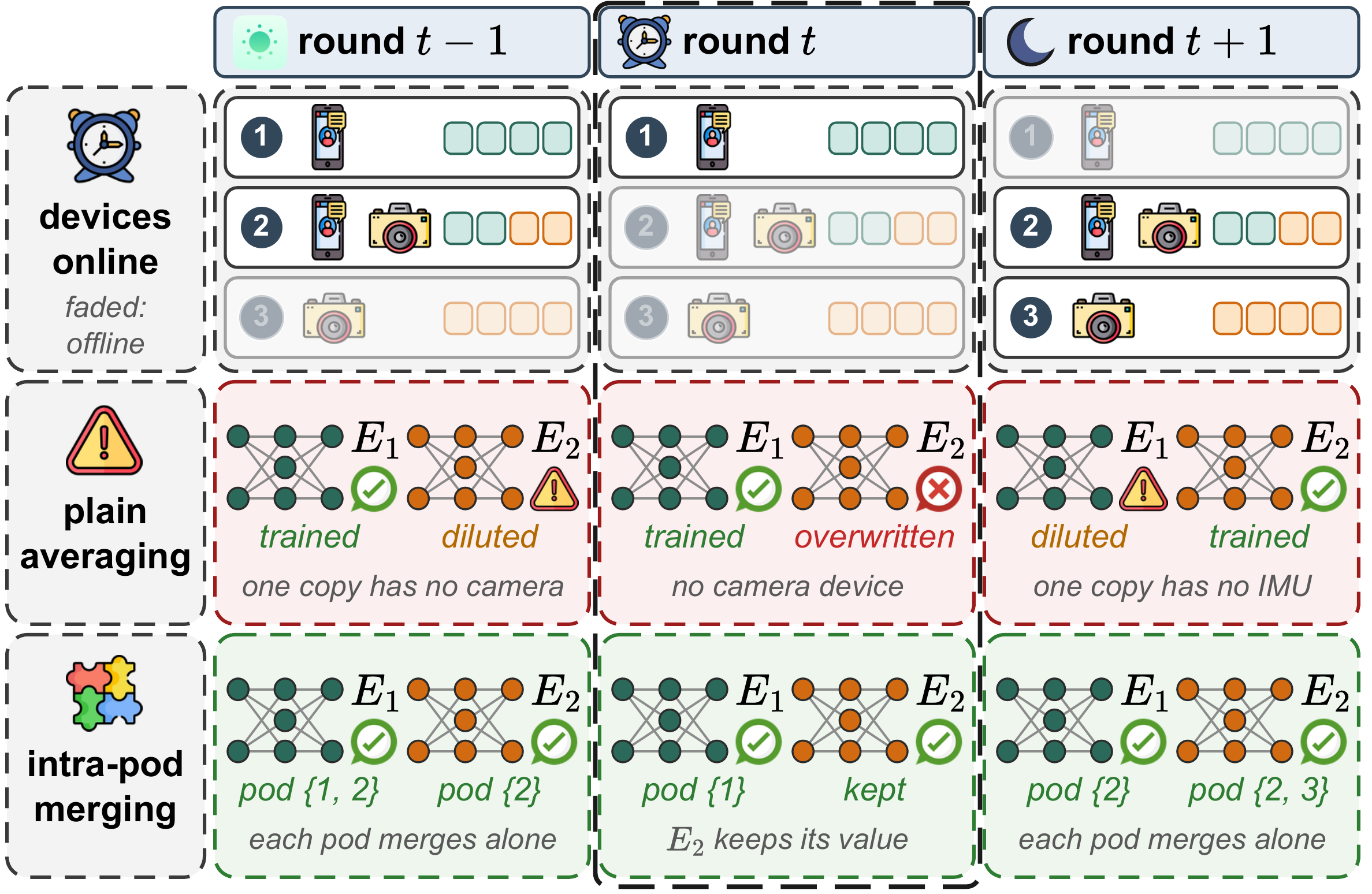}
  \caption{Naive averaging and intra-pod merging over three rounds.}
  \label{fig:intro}
\end{figure}

In real deployments, however, the sensors among the participants change from round to round, for two reasons. (1) Devices carry different sensors, and FL degrades when modalities are missing or mismatched~\cite{zhao2022multimodal,Ding2026ARXIVEASE,Wu2026ARXIVPRISM}. (2) Devices train only when they are idle, charging, and on an unmetered network, so that participation follows periodic patterns~\cite{cho2023convergence,Wu2026ICDCSAlpha,Wu2026ARXIV1}. As such, a sensor held by few devices may be absent for several rounds. Averaging its parameters with those of devices that never trained them then dilutes what its holders have learned (see Fig.~\ref{fig:intro}). Moreover, a single-sensor device never observes how its sensor relates to the others in the first place.

Existing heterogeneity-aware methods fall into two categories, namely, (i) redesigning local training for mismatched sensors and (ii) deciding which devices train. In the first category, Harmony~\cite{ouyang2023harmony} and FedHKD~\cite{wang2024fedhkd} train the sensor encoders and the fusion in separate stages, FedMVD~\cite{gao2025fedmvd} aligns representations across sensor views, and MCARN~\cite{yang2024mcarn} learns a classifier for each sensor with a cross-modal alignment objective. In the second category, CLAMP~\cite{okolo2026clamp} lets slow devices train fewer layers rather than excluding them, and FLPCS-MRF~\cite{wang2025flpcsmrf}, PFHAR~\cite{zhang2026pfhar}, and GraMFedDHAR~\cite{halder2025gramfeddhar} select or train clients for activity recognition. Note that the former are typically evaluated with devices that are always available (or sampled at random), whereas the latter ignore which sensors a device carries or simply assume a complete set.

The sensors that a device carries, however, tell the server which parameters its update can inform. To see this, notice that only the devices holding a sensor train its encoder on real data, while the others merely return copies that never saw it. As such, the parameters of each sensor can be averaged within its ``pod'' (i.e., the set of devices that hold it), and the shared layers over all devices. Note that RELIEF~\cite{wu2026relief} already aggregates the modality blocks of the fusion layer within the devices that hold each modality, and MA-CLAMP~\cite{okolo2026maclamp} aggregates each layer only over the clients that trained it. Furthermore, class prototypes~\cite{tan2022fedproto}, alignment across sensor views~\cite{gao2025fedmvd}, and auxiliary classifiers for each modality~\cite{yang2024mcarn} can carry knowledge to the devices that see a single sensor.

To this end, we design Tier-aware Intra-pod Decoupled Encoders (\method{}). Specifically, unlike RELIEF, \method{} gives each sensor its own encoder and auxiliary classification head, and averages them only within the pod of that sensor, while the fusion layer and the task head remain shared by all devices. In this way, the multi-sensor devices of a pod train the encoders that its single-sensor devices rely on. Class prototypes kept at the server further give every device the class centers of the whole federation, and a scheduler ranks the online devices in terms of speed, data size, and the rarity of their sensor combination.

The contributions of this paper are fourfold. (1) We formulate multimodal FL under periodic availability as maximizing the accuracy reached within a communication budget. (2) We propose \method{}, which trains a classifier on exactly the sensors that each device carries and updates each encoder and its auxiliary head only within its pod. (3) We show in an incremental ablation that the auxiliary heads and intra-pod merging account for most of the gain, that the prototype and cross-sensor losses change accuracy within seed variation, and that score-based selection matches random selection at $K = 20$ but trails it at small $K$ on WEAR. (4) On three activity recognition datasets, \method{} is the most accurate at half its own communication, leads at equal communication on ActionSense and is ahead within seed variation on the other two, and has the second-smallest drop when a sensor is missing.

\section{System Model and Problem Formulation}
\label{sec:system}

\subsection{System Model}
\label{subsec:system-model}

We consider a server and a set $\mathcal{N}$ of $N$ devices that train a global model over rounds indexed by $t$, as illustrated in Fig.~\ref{fig:intro}. The federation uses a set $\mathcal{M}$ of sensors. Device $i$ holds a local dataset $\mathcal{X}_i$ and a subset $\mathcal{M}_i$ of the sensors, with the inputs of the other sensors set to zero. We stipulate the following conventions. Devices with the same sensors are said to form a tier. For each sensor $m$, we call the set of devices that hold it the pod $\mathcal{C}_m$ of sensor $m$. The global model consists of the sensor-specific parameters $\boldsymbol{\theta}_m$ of each sensor $m$ and the shared parameters $\boldsymbol{\theta}^{\textrm{sh}}$.

Device $i$ is online in round $t$ when its availability indicator $a_i[t]$ equals one. Availability is periodic, namely,
\begin{equation}
    a_i[t]\;=\;\mathbb{I}\bigl[(t+\phi_i)\bmod D_i<d_i\bigr],
    \label{eq:avail}
\end{equation}
where $D_i$ is the period, $d_i$ is the number of online rounds per period, and $\phi_i$ is an offset that differs across devices (so that devices with the same pattern are not synchronized). Note that an always-on device has $d_i = D_i$, and that the server knows none of $D_i$, $d_i$, and $\phi_i$. In round $t$, the server observes the set $\mathcal{A}[t]$ of online devices, selects a set $\mathcal{S}[t]$ of at most $K$ of them, and broadcasts the global model $\boldsymbol{\theta}[t]$. Subsequently, each selected device $i$ trains the model on $\mathcal{X}_i$ and returns $\boldsymbol{\theta}_i[t]$. Standard averaging then sets
\begin{equation}
    \boldsymbol{\theta}[t+1]\;=\;\sum_{i\in\mathcal{S}[t]}\frac{|\mathcal{X}_i|}{\sum_{i'\in\mathcal{S}[t]}|\mathcal{X}_{i'}|}\,\boldsymbol{\theta}_i[t].
    \label{eq:fedavg}
\end{equation}

Note that only the members of $\mathcal{C}_m$ provide information about $\boldsymbol{\theta}_m$. Under Eq.~\eqref{eq:fedavg}, however, $\boldsymbol{\theta}_m[t+1]$ also averages the copies $\boldsymbol{\theta}_{i,m}[t]$ returned by the selected devices outside $\mathcal{C}_m$, which have never seen real data for sensor $m$. Worse still, the sensors that the selected devices hold vary from round to round. As such, if no member of $\mathcal{C}_m$ is selected, Eq.~\eqref{eq:fedavg} overwrites $\boldsymbol{\theta}_m$ entirely with copies from devices that do not hold sensor $m$.

\subsection{Problem Formulation}
\label{subsec:problem-formulation}

Let $\Psi[t]$ denote the total model traffic between the server and the devices in both directions up to round $t$ (counting every training stage of multi-stage methods), and let $\mathrm{Acc}[t]$ denote the test accuracy of the global model evaluated in round $t$. The accuracy reached within a communication budget $B$ is then
\begin{equation}
    \mathrm{Acc}@B\;=\;\max\bigl\{\mathrm{Acc}[t]:\Psi[t]\le B\bigr\},
    \label{eq:accB}
\end{equation}
which compares methods at equal cost rather than at their own stopping points. Note that accuracy with every sensor present does not reveal how well the devices with fewer sensors are served. To this end, we also measure (1) the tier-restricted F1, i.e., the macro F1 over the test inputs restricted to the sensors of each less-equipped tier, pooled across those tiers, and (2) the worst-case drop $\Delta$, i.e., the largest accuracy loss when any one sensor is removed from every test input. Now, under the availability of Eq.~\eqref{eq:avail} (which the server does not control), the server chooses the selected sets $\{\mathcal{S}[t]\}$ and an aggregation rule $\mathcal{G}$ by solving
\begin{equation}
\begin{split}
    (\mathrm{P}):\;\;\max_{\{\mathcal{S}[t]\},\,\mathcal{G}}\;\;
    & \mathrm{Acc}@B \\
    \mathrm{s.t.}\;\;
    & (\mathrm{C}_1):\;\mathcal{S}[t]\subseteq\mathcal{A}[t],\;|\mathcal{S}[t]|\le K,\;\forall t, \\
    & (\mathrm{C}_2):\;\boldsymbol{\theta}[t+1]=\mathcal{G}\bigl(\{\boldsymbol{\theta}_i[t]\}_{i\in\mathcal{S}[t]}\bigr),\;\forall t,
\end{split}
\label{eq:problem}
\end{equation}
where $\mathcal{G}$ maps the copies returned in a round to the next global model (Eq.~\eqref{eq:fedavg} being one such rule). A solution to $(\mathrm{P})$ should, in addition, keep the tier-restricted F1 high and $\Delta$ small.

\section{\texorpdfstring{\method{}}{TIDE}: Tier-aware Intra-pod Decoupled Encoders}
\label{sec:method}

\begin{figure*}[t]
  \centering
  \includegraphics[width=0.9\textwidth]{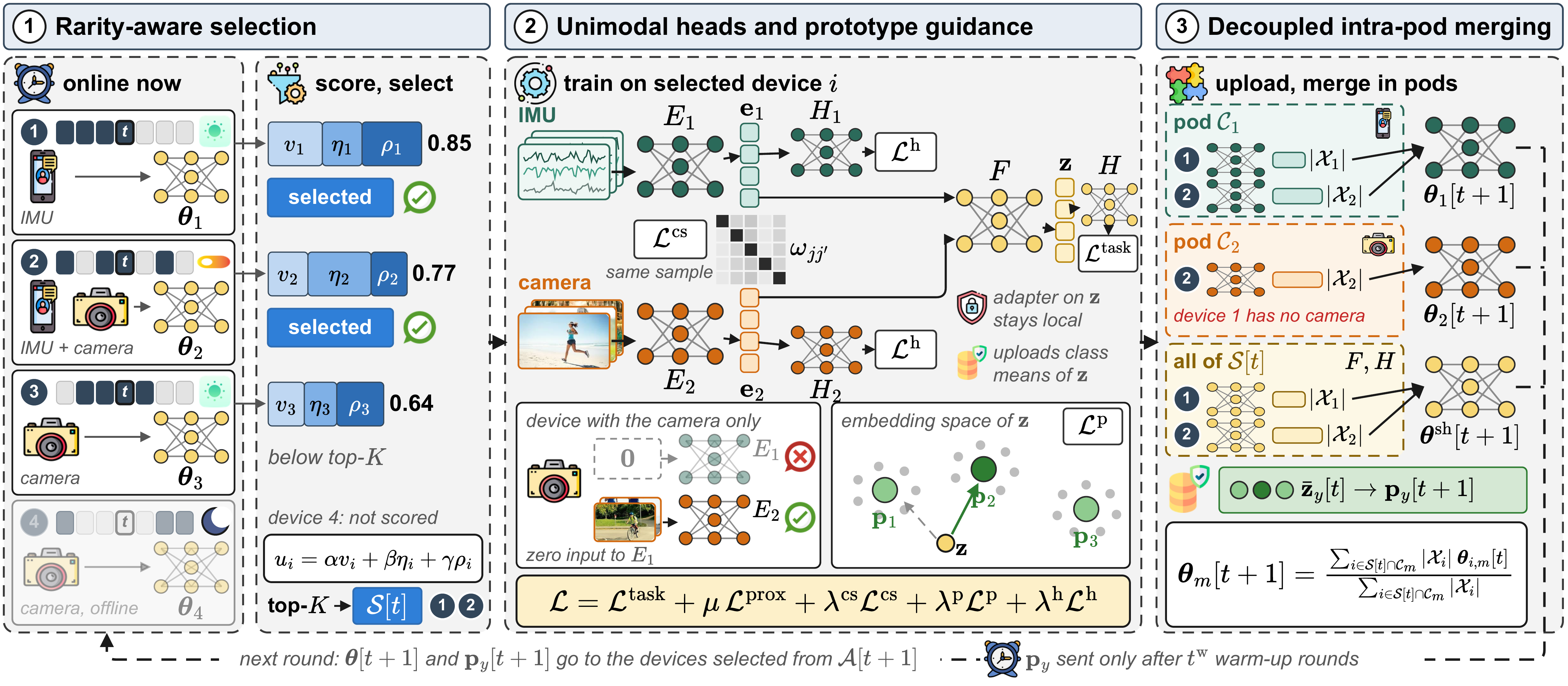}
  \caption{The framework of \method{} with its three modules.}
  \label{fig:framework}
\end{figure*}

\begin{figure}[t]
  \centering
  \includegraphics[width=1\columnwidth]{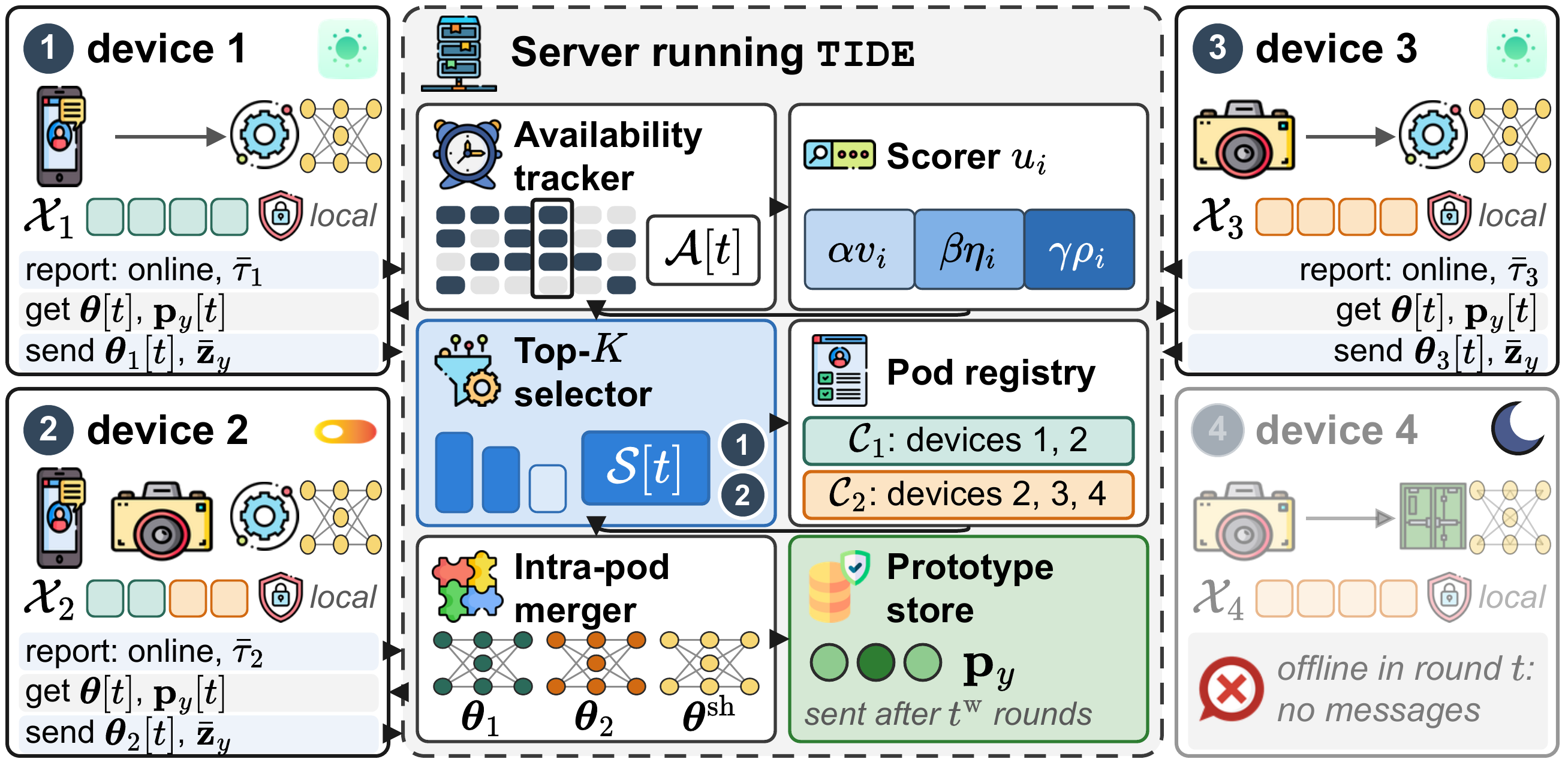}
  \caption{The architecture of \method{} at the server and a device.}
  \label{fig:arch}
\end{figure}

Problem~$(\mathrm{P})$ in Eq.~\eqref{eq:problem} cannot be solved directly, since the server knows none of $D_i$, $d_i$, and $\phi_i$ and hence cannot plan which sensors a round will represent. \method{} therefore fixes $\mathcal{G}$ so that $\boldsymbol{\theta}_m$ depends only on the members of $\mathcal{C}_m$ (Section~\ref{subsec:merge}), which removes the overwriting of Section~\ref{subsec:system-model} in every round, and chooses $\mathcal{S}[t]$ by a heuristic score (Section~\ref{subsec:select}) that Section~\ref{subsec:results} compares with uniform random selection. Figs.~\ref{fig:framework} and~\ref{fig:arch} show the overall framework and where each module runs.

\subsection{Client Selection}
\label{subsec:select}

The purpose of this module is to choose $\mathcal{S}[t]$ subject to $(\mathrm{C}_1)$. Specifically, in every round, \method{} scores each online device $i$ by
\begin{equation}
    u_i\;=\;\alpha\,v_i+\beta\,\eta_i+\gamma\,\rho_i,
    \label{eq:priority}
\end{equation}
where $\alpha$, $\beta$, and $\gamma$ weight the speed score $v_i$, the data score $\eta_i$, and the rarity score $\rho_i$ of the device, respectively, namely,
\begin{equation}
    v_i=\frac{1}{1+\bar{\tau}_i},\quad
    \eta_i=\frac{|\mathcal{X}_i|}{\max_{i'}|\mathcal{X}_{i'}|},\quad
    \rho_i=\frac{\min_{i'}N_{i'}}{N_i}.
    \label{eq:scores}
\end{equation}
Here, $\bar{\tau}_i$ is the mean wall-clock time (in seconds) of the local training of device $i$ over the last 10 rounds, with $v_i$ set to 0.5 before its first round, and $N_i$ is the number of devices whose sensors equal $\mathcal{M}_i$, so that $\rho_i$ is one for the devices with the rarest sensor combination. \method{} then selects the $K$ online devices with the largest $u_i$ (or all of them if fewer than $K$ are online).

\subsection{Unimodal Heads and Prototype-Guided Training}
\label{subsec:local}

Each sensor $m$ has an encoder $E_m$ that maps its input to an embedding $\mathbf{e}_m$, where a sensor that a device lacks (or that is dropped in training) receives an all-zero input. A fusion layer $F$ maps the concatenated embeddings to a fused representation $\mathbf{z}$, from which a task head $H$ predicts the class. In addition, each sensor has an auxiliary head $H_m$ that predicts the class from $\mathbf{e}_m$ alone, so that every encoder learns to be discriminative in its own right. As such, the parameters $\boldsymbol{\theta}_m$ are those of $E_m$ and $H_m$, and $\boldsymbol{\theta}^{\textrm{sh}}$ holds those of $F$ and $H$. Note that each device also keeps a small residual adapter on $\mathbf{z}$ that it never uploads, which holds device-specific adjustments locally (the global model is evaluated without it).

A selected device trains for a fixed number of local epochs on the loss
\begin{equation}
    \mathcal{L}\;=\;\mathcal{L}^{\textrm{task}}+\mu\,\mathcal{L}^{\textrm{prox}}+\lambda^{\textrm{cs}}\mathcal{L}^{\textrm{cs}}+\lambda^{\textrm{p}}\mathcal{L}^{\textrm{p}}+\lambda^{\textrm{h}}\mathcal{L}^{\textrm{h}},
    \label{eq:loss}
\end{equation}
where $\mathcal{L}^{\textrm{task}}$ is the cross-entropy of the task head, $\mathcal{L}^{\textrm{prox}}$ is the squared distance to $\boldsymbol{\theta}[t]$, and $\mathcal{L}^{\textrm{h}}$ is the cross-entropy of the auxiliary heads averaged over the sensors present. The cross-sensor loss $\mathcal{L}^{\textrm{cs}}$ aligns the embeddings of the same sample across sensors. More precisely, for sensors $m$ and $m'$ and a batch of $J$ samples, the scaled similarity between sample $j$ from sensor $m$ and sample $j'$ from sensor $m'$ is
\begin{equation}
    \omega_{jj'}\;=\;\frac{\cos(\mathbf{e}_{m,j},\mathbf{e}_{m',j'})}{\kappa^{\textrm{c}}},
    \label{eq:sim}
\end{equation}
where $\mathbf{e}_{m,j}$ is the embedding of sample $j$ from sensor $m$ and $\kappa^{\textrm{c}}$ is a temperature. The loss of the pair is then
\begin{equation}
\begin{split}
    \mathcal{L}^{\textrm{cs}}_{m,m'}\;=\;-\frac{1}{2J}\sum_{j=1}^{J}\Bigl[&\log\frac{\exp(\omega_{jj})}{\sum_{j'}\exp(\omega_{jj'})}\\
    &+\log\frac{\exp(\omega_{jj})}{\sum_{j'}\exp(\omega_{j'j})}\Bigr],
\end{split}
    \label{eq:cs}
\end{equation}
and $\mathcal{L}^{\textrm{cs}}$ is the mean of $\mathcal{L}^{\textrm{cs}}_{m,m'}$ over all pairs of sensor branches of the model. Note that a zero-input branch has the same embedding for every sample in the batch, and as such, its pairs carry no alignment signal between two real sensors, so Eq.~\eqref{eq:cs} aligns sensors only on devices that hold at least two.

The prototype loss $\mathcal{L}^{\textrm{p}}$ pulls each fused representation toward the server prototype $\mathbf{p}_y[t]$ of its class $y$ and away from the prototypes of the other classes. Specifically, over the classes that already have a prototype,
\begin{equation}
    \mathcal{L}^{\textrm{p}}\;=\;-\frac{1}{J}\sum_{j=1}^{J}\log\frac{\exp\bigl(\cos(\mathbf{z}_j,\mathbf{p}_{y_j}[t])/\kappa^{\textrm{p}}\bigr)}{\sum_{y'}\exp\bigl(\cos(\mathbf{z}_j,\mathbf{p}_{y'}[t])/\kappa^{\textrm{p}}\bigr)},
    \label{eq:proto}
\end{equation}
where $\mathbf{z}_j$ and $y_j$ are the fused representation and the label of sample $j$, respectively, and $\kappa^{\textrm{p}}$ is a temperature. Note that the prototypes are class means of $\mathbf{z}$ over all participants. Hence, Eq.~\eqref{eq:proto} pulls a single-sensor device toward class centers shaped largely by devices with other sensors. In addition, on a multi-sensor device, one random sensor is zeroed in a fraction of the batches (so-called modality dropout), so that the model also sees inputs with a missing sensor.

\begin{table*}[!t]
\centering
\setlength{\tabcolsep}{4pt}
\setlength{\aboverulesep}{0pt}
\setlength{\belowrulesep}{0pt}
\setlength{\extrarowheight}{0pt}
\renewcommand{\arraystretch}{0.88}
\caption{Accuracy (\%), Drop (points), and Comm (GiB) per method and dataset. A dagger marks a run near chance; methods that cannot take a missing sensor are evaluated with it set to zero. Selection rows use $K = 20$ on UCI-HAR, where score-based selection is \methodname{} itself, and 3 and 6 on ActionSense and WEAR. A \swatch{bestcell} cell marks the best and a \swatch{secondcell} cell the second-best value in each band, excluding runs near chance; -- marks no evaluation within the budget.}
\label{tab:main}
\resizebox{\linewidth}{!}{\begin{tabular}{l|cccccc|cccccc|cccccc}
\Xhline{1.2pt}
 & \multicolumn{6}{c|}{\hd{hdrA}{UCI-HAR}} & \multicolumn{6}{c|}{\hd{hdrB}{ActionSense}} & \multicolumn{6}{c}{\hd{hdrC}{WEAR}}\\
\textbf{Method} & Final$\uparrow$ & @B$\uparrow$ & @B/2$\uparrow$ & F1$\uparrow$ & Drop$\downarrow$ & Comm$\downarrow$ & Final$\uparrow$ & @B$\uparrow$ & @B/2$\uparrow$ & F1$\uparrow$ & Drop$\downarrow$ & Comm$\downarrow$ & Final$\uparrow$ & @B$\uparrow$ & @B/2$\uparrow$ & F1$\uparrow$ & Drop$\downarrow$ & Comm$\downarrow$\\
\Xhline{1.2pt}
\multicolumn{19}{l}{\band{(a) Comparison with baselines}}\\
\Xhline{0.6pt}
RELIEF \venue{IoTJ'26} & \best{92.59}\std{0.41} & \second{90.70} & \second{89.29} & \second{0.80} & 21.90 & 280.2 & \second{70.43}\std{1.78} & \second{70.30} & \second{69.42} & 0.61 & 41.58 & 45.7 & 73.23\std{1.04} & 73.23 & 73.23 & 0.55 & 20.86 & 24.7\\
FedMVD \venue{AAAI'25} & 88.99\std{1.48} & 87.23 & 83.56 & 0.68 & 38.41 & 142.9 & 68.57\std{2.47} & 66.49 & 59.35 & 0.61 & 28.82 & 77.9 & \second{79.85}\std{0.74} & \second{79.47} & \second{78.58} & 0.67 & 13.46 & 71.3\\
FLPCS-MRF \venue{Access'25} & 87.39\std{0.98} & 86.77 & 83.78 & 0.39 & 59.30 & 93.8 & 40.16\std{2.53} & 40.16 & 40.16 & 0.26 & 30.76 & \second{17.5} & 69.20\std{0.95} & 69.20 & 69.20 & 0.28 & 52.70 & \best{18.6}\\
MCARN \venue{TPAMI'24} & 82.63\std{1.80} & 48.35 & 39.01 & \best{0.83} & \best{6.19} & 1523.1 & 60.61\std{0.30} & 45.87 & 33.91 & 0.62 & \best{7.84} & 228.1 & 79.15\std{0.18} & 76.54 & 74.46 & \best{0.79} & \best{1.95} & 215.3\\
FedHKD \venue{SenSys'24} & 84.38\std{3.11} & 84.38 & 84.38 & 0.71 & 32.20 & \best{27.7} & 8.78\std{3.46}$^{\dagger}$ & 9.17 & 9.17 & 0.07$^{\dagger}$ & 2.04$^{\dagger}$ & 9.6 & 72.98\std{1.02} & 3.46 & 3.46 & 0.53 & 23.96 & 55.4\\
Harmony \venue{MobiSys'23} & 75.72\std{6.48} & 18.15 & 18.15 & 0.49 & 44.70 & 118.8 & 62.41\std{1.47} & 62.41 & 62.41 & \second{0.64} & 47.16 & \best{7.9} & 70.19\std{1.03} & 70.19 & -- & 0.33 & 55.35 & \second{24.4}\\
PFHAR \venue{TMC'26} & 89.87\std{1.15} & 76.70 & 64.55 & 0.53 & 59.09 & 340.9 & 6.91\std{2.43}$^{\dagger}$ & 6.25 & 5.94 & 0.01$^{\dagger}$ & 2.12$^{\dagger}$ & 41.4 & 61.20\std{0.28} & 61.20 & 61.20 & 0.35 & 19.93 & 28.2\\
GraMFedDHAR \venue{ESWA'27} & 22.42\std{5.90}$^{\dagger}$ & 22.42 & 22.42 & 0.09$^{\dagger}$ & 5.22$^{\dagger}$ & 41.9 & 5.11\std{0.00}$^{\dagger}$ & 5.40 & 5.40 & 0.01$^{\dagger}$ & 0.19$^{\dagger}$ & 16.0 & 55.96\std{2.61} & 47.82 & 44.29 & 0.35 & 13.64 & 159.5\\
\Xhline{0.8pt}
\cellcolor{oursname}\textit{\textbf{\methodname{}}} & \second{91.17}\std{1.03} & \best{90.76} & \best{90.36} & \second{0.80} & \second{21.58} & \second{75.2} & \best{78.06}\std{0.93} & \best{78.06} & \best{76.02} & \best{0.71} & \second{15.47} & 45.2 & \best{80.42}\std{0.77} & \best{80.42} & \best{80.07} & \second{0.69} & \second{10.77} & 41.0\\
\Xhline{0.8pt}
\multicolumn{19}{l}{\band{(b) Ablation of \methodname{}}}\\
\Xhline{0.6pt}
Core & 90.01\std{2.08} & 90.01 & 89.04 & 0.69 & 33.30 & 75.1 & 69.64\std{2.03} & 69.64 & 64.54 & 0.67 & 47.83 & 51.5 & 79.14\std{1.30} & 79.06 & 78.75 & 0.65 & 13.41 & 43.5\\
+ Prototype alignment & 89.84\std{1.19} & 89.54 & 88.49 & 0.71 & 35.01 & 95.6 & 70.70\std{1.00} & 69.46 & 64.01 & \second{0.69} & 50.04 & 66.3 & 79.28\std{1.00} & 79.09 & 78.67 & 0.66 & 12.01 & 54.6\\
+ Participation term & 89.96\std{0.45} & 89.96 & 88.81 & 0.73 & 31.90 & 78.1 & 70.70\std{2.74} & 70.70 & 67.42 & 0.67 & 45.79 & 49.2 & 79.74\std{1.04} & 79.06 & 78.59 & \second{0.68} & 11.83 & 77.8\\
+ Cross-sensor alignment & 90.16\std{0.56} & 89.95 & 89.23 & 0.73 & 32.47 & 79.0 & 71.54\std{0.87} & 70.39 & 65.34 & \second{0.69} & 39.45 & 62.4 & 79.33\std{0.43} & 79.33 & 78.97 & 0.67 & 11.35 & 45.7\\
+ Auxiliary heads & 91.30\std{0.52} & \second{91.30} & 90.14 & 0.79 & \second{21.58} & \second{70.3} & 70.48\std{2.62} & 70.48 & 69.77 & 0.65 & 32.09 & 33.9 & 79.28\std{0.55} & 79.28 & 79.05 & \second{0.68} & \second{10.64} & 44.4\\
+ Intra-pod merging & \best{92.10}\std{0.83} & \best{91.36} & 90.35 & \best{0.82} & \best{16.14} & 112.3 & \best{78.06}\std{0.93} & \best{78.06} & \second{76.02} & \best{0.71} & 15.47 & 45.2 & 80.42\std{0.77} & 80.42 & 80.07 & \best{0.69} & 10.77 & 41.0\\
\Xhline{0.8pt}
\cellcolor{oursname}\textit{\textbf{\methodname{}}} & 91.17\std{1.03} & 90.76 & 90.36 & \second{0.80} & \second{21.58} & 75.2 & \best{78.06}\std{0.93} & \best{78.06} & \second{76.02} & \best{0.71} & 15.47 & 45.2 & 80.42\std{0.77} & 80.42 & 80.07 & \best{0.69} & 10.77 & 41.0\\
\Xhline{0.8pt}
\textit{Evidence-weighted merging} & \second{91.72}\std{0.51} & 91.13 & \best{90.39} & \best{0.82} & \best{16.14} & 111.3 & \second{77.08}\std{0.20} & \second{76.46} & 73.98 & \best{0.71} & \best{12.41} & 63.3 & \second{80.43}\std{1.10} & \second{80.43} & \second{80.24} & \best{0.69} & 10.73 & 39.1\\
\textit{Uniform random selection} & 91.17\std{1.49} & 91.17 & \second{90.38} & 0.76 & 27.61 & \best{63.5} & 76.11\std{1.56} & 76.11 & 75.98 & \second{0.69} & 17.69 & \best{22.4} & \best{80.74}\std{0.31} & \best{80.74} & \best{80.74} & \second{0.68} & 12.04 & \second{21.0}\\
\textit{Score-based selection, same $K$} & 91.17\std{1.03} & 90.76 & 90.36 & \second{0.80} & \second{21.58} & 75.2 & 76.42\std{1.91} & 76.42 & \best{76.33} & \second{0.69} & \second{15.11} & \second{23.5} & 72.52\std{2.09} & 72.52 & 72.52 & 0.60 & \best{8.33} & \best{15.6}\\
\Xhline{1.2pt}
\end{tabular}}
\end{table*}

\subsection{Decoupled Intra-Pod Merging}
\label{subsec:merge}

This module serves as the aggregation rule $\mathcal{G}$ in $(\mathrm{C}_2)$. Rather than Eq.~\eqref{eq:fedavg}, \method{} averages the parameters $\boldsymbol{\theta}_m$ of each sensor only over the selected devices in its pod, weighted by data size, namely,
\begin{equation}
    \boldsymbol{\theta}_m[t+1]\;=\;\frac{\sum_{i\in\mathcal{S}[t]\cap\mathcal{C}_m}|\mathcal{X}_i|\,\boldsymbol{\theta}_{i,m}[t]}{\sum_{i\in\mathcal{S}[t]\cap\mathcal{C}_m}|\mathcal{X}_i|},
    \label{eq:podwise}
\end{equation}
and keeps the previous $\boldsymbol{\theta}_m$ if no selected device holds sensor $m$. The shared parameters $\boldsymbol{\theta}^{\textrm{sh}}$, on the other hand, are averaged over all selected devices as in Eq.~\eqref{eq:fedavg}. In this sense, each encoder is updated only by the devices with real data for it, so the multi-sensor members of a pod train the encoder that its single-sensor members rely on, while the shared layers still pool knowledge from all devices.

Each selected device also returns the mean of $\mathbf{z}$ for each class in its data. The server averages these means into $\bar{\mathbf{z}}_y[t]$ for each class $y$ and updates the associated prototype by
\begin{equation}
    \mathbf{p}_y[t+1]\;=\;0.5\,\mathbf{p}_y[t]+0.5\,\bar{\mathbf{z}}_y[t],
    \label{eq:protoupdate}
\end{equation}
so that no single round can move a prototype abruptly. Note that the prototypes accumulate from the first round, whereas the server sends them to the devices only after a warmup of $t^{\textrm{w}}$ rounds.

\begin{figure*}[!t]
\centering
\subfloat[UCI-HAR.]{\includegraphics[width=0.30\textwidth]{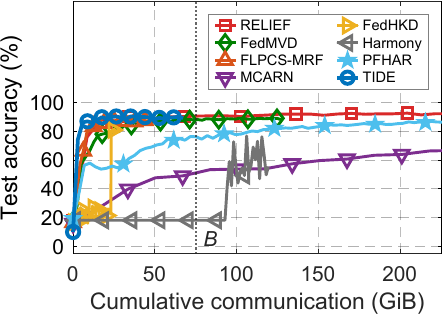}\label{fig:comm_ucihar}}
\hfill
\subfloat[ActionSense.]{\includegraphics[width=0.30\textwidth]{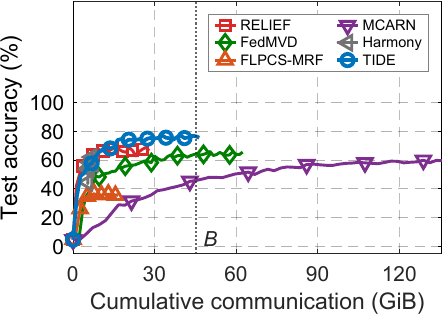}\label{fig:comm_actionsense}}
\hfill
\subfloat[WEAR.]{\includegraphics[width=0.30\textwidth]{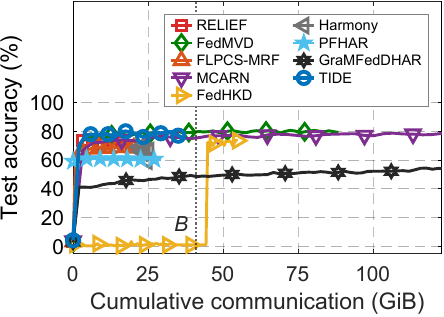}\label{fig:comm_wear}}
\caption{Test accuracy versus cumulative communication in one run of each method, with the budget $B$ dotted.}
\label{fig:comm}
\end{figure*}

\section{Performance Evaluation}
\label{sec:exp}

\subsection{Experimental Setup}
\label{subsec:setup}

We evaluate on three multimodal human activity recognition datasets. From UCI-HAR~\cite{anguita2013public}, every method uses the total acceleration (including gravity) and the angular velocity of a waist-mounted smartphone. Its training split is partitioned into 100 clients by a Dirichlet distribution over labels with concentration 0.5, and its official test split (whose subjects differ) is used for testing. ActionSense~\cite{delpreto2022actionsense} holds eye-tracking, muscle-activity, and body-tracking data from kitchen tasks, whereas WEAR~\cite{bock2024wear} holds inertial data and pre-extracted I3D features of first-person video from outdoor workouts. On these two, each subject is one client, and the windows of each subject are split by time, identically for every method, with the test parts pooled: the last 20\% of the windows of each activity are used for testing, and every training window that shares a sample with a test window is removed. Unlike on UCI-HAR, the test subjects thus also appear in training, but every test window lies later in time. The clients form three tiers by their sensors. On UCI-HAR, 20 clients hold both sensors, 50 the accelerometer only, and 30 the gyroscope only. On ActionSense, three of the nine clients hold all three modalities, three eye tracking and muscle activity, and three muscle activity and body tracking. On WEAR, six of the 18 clients hold both modalities, six inertial data only, and six video only.

Each device follows one of three availability patterns of the form of Eq.~\eqref{eq:avail}, assigned independently of its sensors. Specifically, ``business-hours'' devices are online 40 of every 60 rounds, ``overnight'' devices 20 of every 40, and the remaining devices always, with $\phi_i$ drawn uniformly at random per device. Every method uses the same assignment and selects only among online devices. FedMVD, GraMFedDHAR, and PFHAR use the scores of Eq.~\eqref{eq:priority} plus a participation term of weight 0.15 that rewards recently selected devices, FLPCS-MRF uses its own Pareto-based selection, RELIEF and MCARN train every online client, and Harmony and FedHKD sample uniformly at random. Note that ActionSense and WEAR have fewer clients than $K = 20$ (hence the scores matter only on UCI-HAR), and that all clients share hardware without injected slowdowns (hence the speed score reflects workload rather than hardware).

All methods train from scratch for at most 1000 rounds with five local epochs and batches of 32 samples, evaluate every ten rounds, and stop after ten evaluations without improvement (counted in the final stage of the staged Harmony and FedHKD). Each method uses whichever of the learning rates 0.01 and 0.001 lets it train stably, namely, 0.01 for FedMVD, FLPCS-MRF, and MCARN and 0.001 for the others. \method{} sets $\alpha$, $\beta$, and $\gamma$ to 0.25, 0.35, and 0.25, respectively, $t^{\textrm{w}}$ to 10 rounds, $\mu$ to 0.01, $\lambda^{\textrm{cs}}$ and $\lambda^{\textrm{p}}$ to 0.1, $\lambda^{\textrm{h}}$ to 0.3, and both temperatures $\kappa^{\textrm{c}}$ and $\kappa^{\textrm{p}}$ to 0.3, with a dropout rate of 0.2. PFHAR is reproduced from random initialization without its large-scale pretraining, and hence its results reflect its fine-tuning alone. GraMFedDHAR clips updates at norm 1.0 and adds noise with multiplier 1.0 (or 0.5 on WEAR, where 1.0 diverged). Table~\ref{tab:main} uses the seeds 42, 123, and 2024, and the sensitivity runs use seed 42.

In Table~\ref{tab:main}, Final is the best evaluation of each run on the pooled test data with every sensor present. In Eq.~\eqref{eq:accB}, $B$ is set to the mean total communication of \method{}, i.e., 75.2, 45.2, and 41.0 GiB on UCI-HAR, ActionSense, and WEAR, respectively, which is one point on the curves of Fig.~\ref{fig:comm} and compares every method at the cost that \method{} pays, and we also report the accuracy at $B/2$. Comm is the total communication until termination, converted from rounds at a constant cost per round within each run.\footnote{This conversion is approximate for the staged Harmony and FedHKD.} Note that FLPCS-MRF trains only fully equipped clients, and that our Harmony and FedHKD on UCI-HAR cannot take a missing sensor, so their F1 and Drop set it to zero.

\subsection{Experimental Results}
\label{subsec:results}

\begin{figure*}[!t]
\centering
\subfloat[Selection size.]{\includegraphics[width=0.225\textwidth]{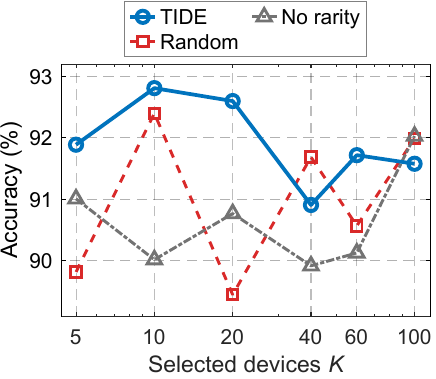}\label{fig:rob_k}}
\hfill
\subfloat[Sensor scarcity.]{\includegraphics[width=0.225\textwidth]{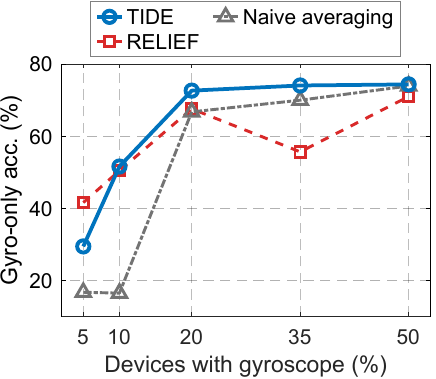}\label{fig:rob_scar}}
\hfill
\subfloat[Availability.]{\includegraphics[width=0.225\textwidth]{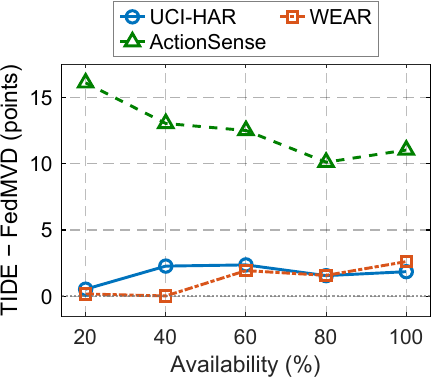}\label{fig:rob_avail}}
\hfill
\subfloat[Training on WEAR.]{\includegraphics[width=0.225\textwidth]{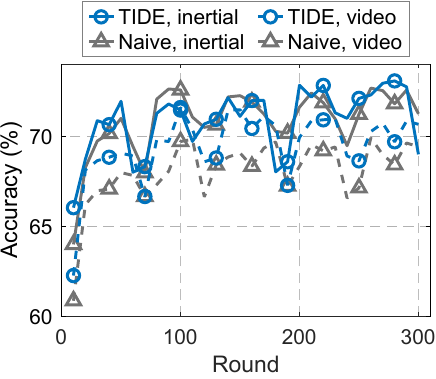}\label{fig:rob_track}}
\caption{Single runs with seed 42. (a) Accuracy versus $K$ and (b) gyroscope-only accuracy versus the share of devices with a gyroscope on UCI-HAR. (c) Accuracy of \method{} minus FedMVD versus the share of each 60-round cycle that every device is online. (d) Per-sensor head accuracy on WEAR.}
\label{fig:rob}
\end{figure*}

In Table~\ref{tab:main}, \method{} is the most accurate method at termination on ActionSense with 78.06\%, i.e., 7.63 points above RELIEF, and on WEAR with 80.42\%, i.e., 0.57 points above FedMVD (within the seed variation) and 1.27 above MCARN. On UCI-HAR, however, RELIEF leads by 1.42 points with 92.59\%. At the budget $B$, \method{} has the highest mean accuracy on all three datasets, by 7.76 points on ActionSense and within the seed variation on UCI-HAR and WEAR. At $B/2$, it is the most accurate on all three, by 1.07 points over RELIEF on UCI-HAR, 6.60 over RELIEF on ActionSense, and 1.49 over FedMVD on WEAR. In our reproduction, GraMFedDHAR did not converge on UCI-HAR (at either noise multiplier) or ActionSense, and PFHAR, without its pretraining, and FedHKD stay near chance on ActionSense. Of the eight baselines, five thus train on every dataset, and the margins above are against these five.

Fig.~\ref{fig:comm} reveals that the method finishing above \method{} on UCI-HAR needs far more communication. Specifically, RELIEF, which trains every online client rather than $K = 20$, uses 280.2 GiB on UCI-HAR (3.7 times that of \method{}) for 1.42 points more, so part of this gap is due to participation size. On ActionSense and WEAR, no baseline finishes above \method{} on average, and MCARN, which trains every online client, uses 5.0 and 5.3 times its communication. Conversely, RELIEF communicates about as much as \method{} on ActionSense and less on WEAR, but stops at 70.43\% and 73.23\%.

\method{} has the highest tier-restricted F1 on ActionSense (0.71) and is within 0.03 of the best on UCI-HAR, but trails MCARN on WEAR (0.69 against 0.79). Its worst-case drop is the second smallest on all three datasets among the methods that train, after MCARN, whose per-sensor classifiers make it the most robust.

The ablation in Table~\ref{tab:main}(b) starts from a ``core'' configuration with the scores of Eq.~\eqref{eq:priority}, the averaging of Eq.~\eqref{eq:fedavg}, the proximal term, and modality dropout. It turns out that the auxiliary heads and intra-pod merging account for most of the gain. Specifically, on top of all alignment losses, the heads add 1.14 points on UCI-HAR but cost 1.06 and 0.05 on ActionSense and WEAR, and intra-pod merging adds a further 0.80, 7.58, and 1.14 points, respectively. Together, they cut the worst-case drop from 32.47 to 16.14 points on UCI-HAR and from 39.45 to 15.47 on ActionSense, while on WEAR it changes from 11.35 to 10.77. Prototype alignment, the participation term, and the cross-sensor loss change the accuracy by at most 1.06 points, i.e., within the seed variation. Note that \method{} omits the participation term, whose removal changes selection only on UCI-HAR, where it lowers the accuracy by 0.93 points (within the seed variation). Weighting the devices of a pod by their head loss (i.e., evidence-weighted merging) gives no consistent gain either. Changing one of $\lambda^{\textrm{h}} \in \{0.1, 0.5, 1.0\}$, $\lambda^{\textrm{p}}, \lambda^{\textrm{cs}} \in \{0.05, 0.3\}$, or $t^{\textrm{w}} \in \{0, 50\}$ at a time keeps the accuracy within 1.74, 2.40, and 0.79 points of the default mean on UCI-HAR, ActionSense, and WEAR, respectively (within the seed variation on UCI-HAR and WEAR), and the worst-case drop at 14.5--18.8, 12.4--15.4, and 8.9--12.3 points, except that $\lambda^{\textrm{p}} = 0.3$ raises it to 36.3 on WEAR.

On UCI-HAR, the data and rarity scores do not change during training, and as such, the 20 clients with both sensors receive 50\% to 60\% of all selections, while 64 to 67 of the 100 clients are never selected. Interestingly, random selection gives the same accuracy within the seed variation, but a lower tier-restricted F1 (0.76 against 0.80) and a larger worst-case drop (27.6 against 21.6), since under score-based selection the encoders of the less-equipped devices are trained within their pods by the clients that hold both sensors. With few selected devices, however, the fixed scores keep choosing the same ones. Specifically, with $K = 6$ on WEAR, score-based selection trails random selection by 8.22 points, whereas with $K = 3$ on ActionSense the two give the same accuracy within the seed variation, and random selection stays within 1.95 points of full participation with 50\% and 51\% of its communication.

Finally, Fig.~\ref{fig:rob} varies one condition at a time with seed 42. On UCI-HAR, all three selection rules reach 89.45\% to 92.81\% for $K$ from 5 to 100, and no rule is consistently ahead. With a gyroscope on only 10\% of devices, naive averaging reaches a gyroscope-only accuracy of 16.42\% against 51.58\% for \method{}, whereas RELIEF is the most accurate at 5\%. When every device is online for the same share of each cycle, \method{} stays within 1.7 and 2.4 points of its best accuracy on UCI-HAR and ActionSense down to 20\%, and above FedMVD at every level on all three datasets (by 0.54 to 2.37, 10.11 to 16.09, and 0.03 to 2.62 points, respectively), although the two are level on WEAR at 20\% and 40\%. Also on WEAR, the head of each sensor ends within 2.25 points of its accuracy under naive averaging.

\section{Conclusion}
\label{sec:conclusion}

In this paper, we have studied multimodal federated learning for IoT devices that carry different sensors and come online in periodic patterns. To this end, we proposed \method{}, which merges the encoder and auxiliary head of each sensor only within its pod (i.e., the devices that hold the sensor), while the fusion layer and task head remain shared by all devices. On three activity recognition datasets, \method{} is the most accurate at half its own communication, leads at equal communication on ActionSense, is ahead within seed variation on UCI-HAR and WEAR, and has the second-smallest worst-case drop when a sensor is missing, after MCARN. Note that its selection scores do not change during training. As such, when few devices are selected, it unfortunately trails random selection on WEAR. Adapting the scores during training and evaluating on devices with different hardware are left for future work.

\balance

\end{document}